\documentclass[12pt,a4paper,dvips]{article}
\usepackage{epsfig,wrapfig,times,mathptm} 
\renewcommand{\section}[1]{\vspace{6pt} \noindent\mbox{#1} \newline \noindent}
\renewcommand{\subsection}[1]{\vspace{6pt} \noindent\mbox{\underline{#1}} 
\newline \noindent}
\renewcommand{\subsubsection}[1]{\vspace{6pt} \noindent\mbox{\underline{#1}}
\noindent}

\newfont{\sansb}{cmssbx10}
\newfont{\sans}{cmss10}

\begin{document}
{\small ICRC97 , Pre-conference Workshop  
\vspace{-24pt}\\}     %The session code
{\center \LARGE POSSIBLE EFFECTS OF LORENTZ SYMMETRY VIOLATION 
ON THE INTERACTION PROPERTIES OF VERY HIGH-ENERGY COSMIC RAYS
\vspace{6pt}\\}
L. Gonzalez-Mestres$^{1,2}$ \vspace{6pt}\\
{\it $^1$Laboratoire de Physique Corpusculaire, Coll\`ege de France, 
75231 Paris Cedex 05 , France\\
$^2$Laboratoire d'Annecy-le-Vieux de Physique des Particules, 74941 
Annecy-le-Vieux Cedex,
France
 \vspace{-12pt}\\}
{\center ABSTRACT\\}
Special relativity has been tested at low energy with great accuracy, but
these results cannot be extrapolated to the very high-energy region.
Introducing a critical distance scale, $a$ , 
below $10^{-25}~ cm$ (the wavelength
scale of the highest-energy observed cosmic rays) allows to consider models,
compatible with standard tests of special relativity, where a small
violation of Lorentz symmetry ($a$ can, for instance, be the Planck length
$\approx 10^{-33}~cm$)
produces dramatic effects on the interaction properties of very high-energy 
particles. Lorentz symmetry violation may potentially solve all 
the basic problems
raised by the highest-energy cosmic rays (origin 
and energy, propagation...).
Furthermore, superluminal sectors of matter may exist and 
release very high-energy ordinary particles or directly
produce very high-energy 
cosmic-ray events with unambiguous signatures in very large detectors. 
We discuss 
these phenomena, as well as the cosmic-ray energy
range (well below the energy scale associated to the fundamental length)
and experiments where they could be detected and studied. 

\setlength{\parindent}{1cm}
\section{LORENTZ SYMMETRY VIOLATION BY NONLOCAL DYNAMICS}
It is a common prejudice that Lorentz invariance may cease to be valid at
Planck energy, but the consequences of this possibility for physics at lower
energy scales remain basically unexplored. 
Contrary to superficial appearance, such a phenomenon would indeed  
play a crucial role at energies well below Planck scale. In particular,
important observable effects can be predicted at the energy scale of the
highest-energy cosmic rays.
In previous papers (Gonzalez-Mestres, 1997a and 1997b), we suggested
that, as a consequence of nonlocal dynamics at Planck scale or at some other
fundamental length scale, Lorentz symmetry violation can result in a
modification of the equation relating energy and momentum which
would write in the vacuum rest frame:
\equation
E~~=~~(2\pi )^{-1}~h~c~a^{-1}~e~(k~a)
\endequation
where $E$ is the energy of the particle,
$h$ the Planck constant, $c$ the speed of light,
$a$ the fundamental length scale, $k$  
the wave vector modulus and
$[e~(k~a)]^2$ is a convex
function of $(k~a)^2$ obtained from nonlocal vacuum dynamics.
Rather generally, we find
that, at wave vector scales below the inverse of the fundamental length scale,
Lorentz symmetry violation in relativistic kinematics can be parameterized
writing:
\equation
e~(k~a)~~\simeq ~~[(k~a)^2~-~\alpha ~(k~a)^4~+~(2\pi ~a)^2~h^{-2}~m^2~c^2]^{1/2}
\endequation
where $\alpha $ is a positive
constant between $10^{-1}$ and $10^{-2}$ . At high energy, we can write:
\equation
e~(k~a)~~\simeq ~~k~a~[1~-~\alpha ~(k~a)^2/2]~
+~2~\pi ^2~h^{-2}~k^{-1}~a~m^2~c^2
\endequation
and, in any case, we expect observable kinematical effects when the term
$\alpha (ka)^3/2$ becomes as large as the term
$2~\pi ^2~h^{-2}~k^{-1}~a~m^2~c^2$ .
For a proton
at $E~\approx ~ 10^{20}~eV$ and with $a~\approx ~10^{-33}~cm$ , one would have:
\equation
\alpha ~(k~a)^2/2~~\approx ~~10^{-18} ~~\gg ~~2~\pi ^2~h^{-2}~k^{-2}~m^2~c^2~~
\approx 10^{-22}
\endequation
so that, although $\alpha (ka)^3/2$ is indeed very small as compared to the
value of $e~(k~a)$ , the term $2~\pi ^2~h^{-2}~k^{-1}~a~m^2~c^2$ represents
an even smaller fraction of this quantity. We therefore expect corrections
to relativistic kinematics to play a crucial role at the highest
cosmic ray energies, well below the energy scale associated to the inverse
of the fundamental length. In what follows, we make the crucial assumption
that the earth is moving
slowly with respect to the vacuum rest frame.

\section{CONSEQUENCES OF THE NEW KINEMATICS}
Particles at the wavelentgth scale of the highest-energy cosmic rays would 
be sensitive to the new kinematics. If  $\alpha $ were negative, all particles
would become unstable at these energies, more precisely at energies above
$E~\approx ~(2\pi )^{-1/2}~\alpha ^{-1/4}~(m~c^3~h~a^{-1})^{1/2}$ .
For instance, if $c$ and $\alpha $ had universal values for all particles
and $\alpha $ were negative, and taking $a~\approx ~10^{-33}~cm$ ,
a proton with energy above $\approx ~10^{19}~eV$ would decay into a proton
of lower energy and one or several photons, or into a neutron or a proton
plus pions, or by emitting particle-antiparticle pairs. 
Similar considerations would apply to electrons above $\approx ~10^{17}~eV$ .
The unstability
of very high-energy particles would not disappear if the value of $\alpha $ ,
although negative, were not universal. Any very high-energy particle
with $\alpha <~0$ 
could always decay by emitting its own particle-antiparticle pairs,
even if $c$ were not universal (in this Section, we naturally
assume the breaking of the universality of $c$ to be a rather small effect).
With $\alpha ~=~0$ , some very high-energy particles would be unstable if 
$c$ were not universal whereas others would remain stable
(Coleman and Glashow, 1997); but, if $\alpha ~\neq ~0$ , 
new mechanisms arise (Gonzalez-Mestres, 1997a and 1997b) which compete with
that considered by Coleman and Glashow and, due the wavelength dependence of
nonlocal effects, dominate over the Coleman-Glashow (CG) mechanism at high 
enough energy. 
With $\alpha ~>~0$ for all particles, the effect of the term 
$-~\alpha ~(k~a)^2/2$ dominates over the CG mechanism
at very high energy, even if the value
of $\alpha $ is not universal.

If $c$ and $\alpha $ have universal values, and $\alpha $ is positive, 
and taking the Planck length 
to be the fundamental distance scale, the following new effects 
arise:

a) The Greisen-Zatsepin-Kuzmin (GZK) cutoff on very high-energy cosmic
protons and nuclei (Greisen, 1966;
Zatsepin and Kuzmin, 1966) does no longer apply (Gonzalez-Mestres, 
1997a and 1997b). Very high-energy cosmic rays originating
from most of the presently
observable Universe can reach the earth and generate the highest-energy 
detected events.

b) Unstable particles with at least two massive particles in the final state
of all their decay channels become stable at very high energy 
(Gonzalez-Mestres, 1997a and 1997b). 
In any case, unstable particles live longer than naively expected with exact
Lorentz invariance and, at high enough energy,
the effect becomes much stronger than previously estimated for nonlocal models
(Anchordoqui, Dova, G\'omez Dumm
and Lacentre, 1997)
ignoring the small violation of relativistic kinematics.

c) The allowed final-state phase space of two-body collisions is seriously
modified at very high energy, especially when, in the vacuum rest frame
where expressions (1) - (3) apply, 
a very high-energy particle 
collides with a low-energy target (Gonzalez-Mestres, 1997c). Energy
conservation reduces the final-state
phase space at very high energy and can lead to a sharp fall of cross sections
starting at incoming-particle wave vectors well below the
inverse of the fundamental length.
 
d) In astrophysical processes,
the new kinematics may inhibit phenomena such as GZK-like cutoffs,
photodisintegration of nuclei, decays,
radiation emission under external forces, momentum loss
(which at very high energy does not imply deceleration) through collisions,
production of lower-energy secondaries... {\bf potentially solving all 
the basic
problems raised by the highest-energy cosmic rays}
(Gonzalez-Mestres, 1997c). Due to the fall of 
cross sections, energy losses become much weaker than expected with 
relativistic kinematics and astrophysical particles can be pushed to much 
higher energies; similarly, they will be able to propagate to 
much longuer astrophysical distances, and many more possible sources 
(in practically all the presently observable Universe) can be
considered for very high-energy cosmic rays reaching the 
earth; as particle lifetimes are
much longer, new possibilities arise for the nature of these cosmic rays.
The same considerations apply to nuclei.

e) If the new kinematics can explain the existence of $\approx 10^{20}~eV$ 
events, it also predicts that, above some higher energy
scale (around $\approx 10^{22}~eV$ for $a~\approx ~10^{-33}~cm$), the fall of
cross sections will prevent the cosmic ray from depositing most of its 
energy in the atmosphere (Gonzalez-Mestres, 1997c).   
Such extremely-high energy particles will produce atypical events of
apparently much lower energy. New analysis
of data and experimental designs are required to explore 
this possibility.

To be complete, our discussion must also consider situations where $\alpha $ is
positive, but its value is not universal. This could be the case, for instance, if the composite character of nucleons and nuclei prevents them from being
described as "elementary" 
single particles when deriving expressions like (1) - (3). Due to
the loss of Lorentz invariance, it is far from obvious how to extrapolate
to Planck scale the low-energy bound-state structure: the rest structure
of the composite object
will depend on its speed with respect to the vacuum rest frame.
The value of $\alpha $ may depend on this dynamics.

In a parton-like picture,
we could be led to $\alpha ~\approx ~0.05$ for quarks, leptons
and gauge bosons, $\approx ~0.01$ for mesons,
$\approx ~0.005$ for nucleons and $\approx ~0.05~(3A)^{-2}$
for a nucleus with $A$ nucleons. Such a kinematics implies upper
bounds on the allowed energy of a stable nucleon or nucleus (above some energy,
spontaneous emission of photons or electron-positron pairs is kinematically
allowed). 
In this case, with $a~\approx ~10^{-33}~cm$ 
and assuming a universal value of $c$ , cosmic ray events above 
$\approx 10^{19}~eV$ cannot be protons, and above $\approx 10^{20}~eV$ they 
cannot be nuclei. It may be tempting to try to tune the model in such a
way that the maximum allowed energy for cosmic nuclei coincides with that
of the most energetic observed cosmic ray event (around $3.10^{20}~eV$).
Among the parameters of the fit would be the fundamental length $a$ and,
for each particle, the values of $c$ and $\alpha $ .
A more unconventional approach would be to attribute events above 
$\approx 10^{20}~eV$ to other kinds of cosmic particles. 
For instance, at $E~\approx 10^{18}~eV$ and with $a~\approx ~10^{-33}~cm$ , 
a muon would have a lifetime $~\approx 10^4~s$ 
according to standard relativity; but, in the model we are considering,  
Lorentz symmetry violation would reduce the phase space of its decay 
products at this energy scale, and more and more strongly as energy increases
(Gonzalez-Mestres, 1997a and 1997b).
It is not excluded that, in very high astrophysical magnetic
fields, muons can live long enough to be 
pushed to ultra-high energies. Becoming very long-lived with the help
of Lorentz symmetry violation
and keeping their energy with the help of the fall of electromagnetic 
cross sections (Gonzalez-Mestres, 1997c), 
they may subsequently reach the earth. When studying the event shape 
in detectors, it should also
be kept in mind that, at such energies never reached by accelerators, new
interaction properties may develop.

In general, if $\alpha $ is positive but does not have a
universal value, precise numerical studies are required for each specific
scenario, but candidates to the highest-energy cosmic-ray events seem to
exist in most cases. It is remarkable that future experiments devoted to 
ultra-high energy cosmic rays will indeed be able to test models involving
Lorentz symmetry violation at Planck scale and discriminate between different
scenarios, using the shape and spectra of cosmic-ray events. 

\section {SUPERLUMINAL PARTICLES}
{\bf Superbradyons}, i.e. superluminal particles with positive mass and energy
and critical speed in vacuum $c_i~\gg ~c$ ($c$ = speed of light), have been 
considered in previous papers (Gonzalez-Mestres, 1996 and 1997d). They can 
be associated to new dynamical sectors where, for the $i$-th sector, a 
Lorentz symmetry exists with critical speed parameter $c_i$ . Mixing between
different dynamical sectors would break these Lorentz invariances, including
the "ordinary" one (i.e. that of standard relativity,
with critical speed parameter equal to $c$). Superluminal particles can have
very large rest energies due to the relation $E~=~mc_i^2$ 
(Gonzalez-Mestres, 1996 and 1997d), and produce very high-energy "ordinary"
cosmic rays through astrophysical processes, decays, collisions or "Cherenkov"
radiation in vacuum (spontaneous emission of particles with lower critical 
speed). They can also reach the earth and produce inelastic events in very
large-volume detectors: then, due to energy and momentum conservation
and to the relation $E~\simeq ~c_i~p~\gg ~cp$ , an incoming superluminal
particle of energy $E$ , momentum $p$ and critical speed in vacuum $c_i$
would originate two very high-energy showers with almost
exactly equal energy and momentum, and almost exactly "back-to-back" 
(Gonzalez-Mestres, 1996 and 1997d). The situation is similar for
spontaneous "Cherenkov" emission. Underground or underwater very 
large-volume detectors would be able to unambiguously identify such events,
even at extremely low rate. But, for very high-energy events, a very
well-suited "large-volume" detector would also be the AUGER
observatory (AUGER Collaboration, 1997), which will most likely offer the 
largest available target (the atmosphere over
a uniquely large surface) and where unconventional events will develop
atypical cascade development profiles, again providing 
unambiguous signatures. 
 
\section{CONCLUDING REMARKS}
Future experiments devoted to ultra-high energy cosmic rays will provide
unique tests of 
possible Lorentz symmetry violation at scales unaccessible to standard
tests, and be sensitive to possible departures from the 
Poincar\'e relativity principle (Poincar\'e, 1905) at very small distances. 
At least two kinds of phenomena violating this principle deserve
serious consideration: a) nonlocal effects at Planck scale or at some other
fundamental length scale; b) interaction with superluminal sectors of matter.
In the first case, the behaviour of "ordinary"
particles will be modified. In the second case, we can try to directly discover
the superluminal particles.
Contrary to
general prejudice, a violation of Lorentz symmetry at distance scale $\approx
~a$ does not require energies close to $\approx (2\pi )^{-1}~h~c~a^{-1}$ to
produce observable effects at leading level.
Assuming $c$ and $\alpha $ ($\alpha ~>~0$) to have universal values, 
cosmic ray data suggest $a$ to be in the range $10^{-35}~cm~<~a~<~10^{-30}~cm$
(Gonzalez-Mestres, 1997c). Superluminal particles can 
produce very high-energy events in very large-surface or very large-volume 
detectors: future experiments should consider their possible
existence. Very high-energy "ordinary" cosmic rays may also have been released
by cosmic superluminal particles. Very high-energy
interactions (as measured
in the vacuum rest frame)
in astrophysical processes, as well as in cosmic-ray propagation
and in cosmic-ray detectors, are the key to possibly uncover and identify
this new physics.  

\section{REFERENCES}
\setlength{\parindent}{-5mm}
\begin{list}{}{\topsep 0pt \partopsep 0pt \itemsep 0pt \leftmargin 5mm
\parsep 0pt \itemindent -5mm}
\vspace{-15pt}
\item
Anchordoqui, L., Dova, M.T., G\'omez Dumm, D. and Lacentre, P., 
{\it Zeitschrift f\"{u}r Physik C} 73 , 465 (1997).
\item
AUGER Collaboration, "The Pierre Auger Observatory Design Report" (1997).
\item
Coleman, S. and Glashow, S.L., "Cosmic Ray and Neutrino Tests of Special 
Relativity", paper hep-ph/9703240 of LANL (Los Alamos) 
electronic archive (1997).
\item
Gonzalez-Mestres, L., "Physical and Cosmological Implications of a Possible
Class of Particles Able to Travel Faster than Light", contribution to the
28$^{th}$ International Conference on High-Energy Physics, Warsaw July 1996 .
Paper hep-ph/9610474 of LANL electronic archive (1996).
\item
Gonzalez-Mestres, L., "Vacuum Structure, Lorentz Symmetry and Superluminal
Particles", paper physics/9704017 of LANL  
electronic archive (1997a).
\item
Gonzalez-Mestres, L., "Absence of Greisen-Zatsepin-Kuzmin Cutoff
and Stability of Unstable Particles at Very High Energy,
as a Consequence of Lorentz Symmetry Violation", contribution HE 1.2.36 
to the 25-th International Cosmic Ray Conference. Paper
physics/9705031 of LANL electronic archive (1997b).
\item
Gonzalez-Mestres, L., "Lorentz Symmetry Violation and Very High-Energy Cross
Sections", contribution to the International Conference on Relativistic 
Physics and some of its Applications, Athens June 1997 . Paper physics/9706022
of LANL
electronic archive (1997c).
\item
Gonzalez-Mestres, L., "Superluminal Particles and High-Energy Cosmic Rays",
contribution HE 1.2.35 to the 25-th International Cosmic Ray Conference. 
Paper physics/9705032  of LANL electronic archive (1997d).
\item
Greisen, K., {\it Phys. Rev. Lett.} 16 , 748 (1966).
\item
Kirzhnits, D.A., and Chechin, V.A., {\it Soviet Journal of Nuclear Physics}, 
15 , 585 (1972).
\item
Poincar\'e, H., Speech at the St. Louis International Exposition of 1904 ,
{\it The Monist 15} , 1 (1905).
\item
Zatsepin, G.T. and 
Kuzmin, V.A., {\it Pisma Zh. Eksp. Teor. Fiz.} 4 , 114 (1966).

\end{list}

\end{document}